\documentclass[runningheads]{llncs}
\AtBeginDocument{%
  \providecommand\BibTeX{{%
    \normalfont B\kern-0.5em{\scshape i\kern-0.25em b}\kern-0.8em\TeX}}}
\usepackage{siunitx}
\usepackage{amsmath,amsfonts}
\usepackage{algorithm}
\usepackage{xcolor}
\usepackage{moreverb}
\usepackage{tabularx}
\usepackage{threeparttable} 
\usepackage{multirow} 
\usepackage{rotating} 
\usepackage{bigstrut} 
\usepackage[utf8]{inputenc}
\usepackage{paralist}
\usepackage{lipsum}
\usepackage{soul}
\usepackage[noend]{algpseudocode}
\usepackage{subfig} 
\usepackage{caption}
\usepackage{lipsum}
\usepackage{indentfirst}
\usepackage[labelsep=period]{caption}
\usepackage{todonotes}
\newboolean{showcomments}
\setboolean{showcomments}{true}         

\ifthenelse{\boolean{showcomments}}
{\newcommand{\nb}[2]{
		\fbox{\bfseries\sffamily\scriptsize#1}
		{\sf\small$\blacktriangleright$\textit{\textcolor{red}{#2}}$\blacktriangleleft$}
	}
}
{\newcommand{\nb}[2]{}
	
}

\begin{document}

\title{
Towards Robust IoT Defense: Comparative Statistics of Attack Detection in Resource-Constrained Scenarios}



\author{Zainab Alwaisi\inst{1}
\and
Simone Soderi\inst{1,2}
}
\authorrunning{Z. Alwaisi et al.}

\institute{IMT School for Advanced Studies, Lucca, Italy\\
\email{\{zainab.alwaisi, simone.soderi\}@imtlucca.it} \and
CINI Cybersecurity Laboratory, Roma, Italy 
}

%
\maketitle

\begin{abstract}
Resource constraints pose a significant cybersecurity threat to IoT smart devices, making them vulnerable to various attacks, including those targeting energy and memory. This study underscores the need for innovative security measures due to resource-related incidents in smart devices. 
In this paper, we conduct an extensive statistical analysis of cyberattack detection algorithms under resource constraints to identify the most efficient one. Our research involves a comparative analysis of various algorithms, including those from our previous work. We specifically compare a lightweight algorithm for detecting resource-constrained cyberattacks with another designed for the same purpose. Notably, the latter employs TinyML for detection.
In addition to the comprehensive evaluation of the proposed algorithms, we introduced a novel detection method for resource-constrained attacks. This method involves analyzing protocol data and categorizing the final data packet as normal or attacked. Subsequently, the attacked data is further analyzed in terms of the memory and energy consumption of the devices to determine whether it's an energy or memory attack or another form of malicious activity.
We compare the suggested algorithm performance using four evaluation metrics: accuracy, probability of detection, probability of false alarm, and probability of misdetection. Notably, the proposed dynamic techniques dynamically select the classifier with the best results for detecting attacks, ensuring optimal performance even within resource-constrained IoT environments. The results indicate that the proposed algorithms outperform the existing works with accuracy for algorithms with TinyML and without TinyML of $99.3$\%, $98.2$\%, a probability of detection of $99.4$\%, $97.3$\%, a probability of false alarm of $1.23$\%, $1.64$\%, a probability of misdetection of $1.64$\%, $1.46$ respectively.  In contrast, the accuracy of the novel detection mechanism exceeds $99.5\%$ for RF and $97\%$ for SVM.

\end{abstract}


\section{Introduction}
The Internet of Things (IoT) facilitates the connection of numerous devices, enabling data exchange and updates for various applications across smart city infrastructure, smart homes, smart vehicles, advanced healthcare systems, and other intelligent environments. The IoT network requires swift communication with low latency to avoid delays and high throughput to accommodate the vast number of devices through wireless or wired connections. IoT devices can connect directly to the Internet using technologies like cellular connections or Wi-Fi or indirectly via specific IoT protocols. In the latter case, the devices initially communicate with a base station or a gateway, and then the gateway acts as a bridge to connect to the internet~\cite{Khor2021}. The complexity of IoT models is steadily increasing~\cite{pahl2018all,Pahl18}. People are becoming increasingly accustomed to data-driven infrastructure, driving the research of more Machine Learning-based applications in conjunction with IoT. IoT and Machine Learning (ML) techniques are currently being applied in virtually every facet of human life. IoT devices utilize wireless communication, making them susceptible to potential security breaches~\cite{Liu18}. Unlike typical communication attacks in local networks, which are often limited to local nodes or small local domains, attacks on IoT systems can extend over a broader area, causing significant disruptions to IoT installations~\cite{pajouh2016two}.
As energy efficiency demands in IoT applications rise, the need for low latency, enhanced reliability, and strengthened security also increases. Particularly crucial in applications like smart cities, digital healthcare, and Industry 4.0~ \cite{Kamaldeep2020}. Integrating edge and fog computing into conventional IoT networks provides substantial advantages by placing computational resources closer to devices and data sources. Incorporating Artificial Intelligence (AI) and other advanced technologies enables automated and expedited data processing, analysis, and decision-making. In prior research~\cite{BICT2023,Fabulous2023,INISCOM2023,alwaisi2023optimized}, we extensively investigated resource-constrained attacks on smart devices, developing lightweight detection mechanisms for timely monitoring. Emphasizing energy and memory attack analysis, these solutions focus on detecting and monitoring attacks effectively.

Subsequently, recognizing the potential of TinyML as an additional approach to detect resource-constrained attacks on smart devices, we explored the integration of TinyML as an innovative solution to address resource constraints more effectively. Leveraging TinyML in this context significantly enhances the cybersecurity posture of smart devices. It enables real-time attack detection and the capacity to predict future attacks through ML, which learns from resource-based attacks in their nascent stages.
The statistical analysis to show the best detection mechanism to detect resource constraint attacks is proposed to address the challenges arising from the increased utilization of IoT devices, which generate substantial volumes of data overwhelming security providers. This statical analysis advocates for cybersecurity providers to effectively detect resource constraints and cyber attack problems and understand the best mechanisms to tackle this security issue. Advanced analytics and ML algorithms play a crucial role in achieving this objective, enabling the identification of patterns and trends in the data and prioritization of the most pertinent information for clinicians. By leveraging the potential of IoT data through these approaches, cybersecurity providers can enhance the security outcomes of IoT smart devices and elevate the overall quality of detecting such attacks. 

Moreover, this paper proposes a resource constraints attack detection method based on ML techniques. Based on the previous research papers~\cite{BICT2023,Fabulous2023,INISCOM2023,alwaisi2023optimized}, through the analysis of the principle of resource constraints attack, the three common attack packets obtained by operating the energy and memory attacks tool are grouped in the feature extraction stage. The characteristics of attack flow are obtained through the analysis of normal flow data. The attack traffic characteristics obtained in the model detection phase are trained in the training model based on the Random Forest (RF) algorithm. Finally, the test model is validated by the resource constraints attack, and the  Support Vector Machines (SVM) method in ML is compared in terms of detection accuracy. The results show that the resource constraints attack detection method based on ML proposed in this paper has a reasonable detection rate for the current popular energy and memory attacks. In addition, this method involves analyzing protocol data and categorizing the final data packet as normal or attacked. Subsequently, the attacked data is further analyzed in terms of the memory and energy consumption of the devices to determine whether it's an energy or memory attack or another form of malicious activity. To specify the type of attack, we used ML techniques to compare the fetched attacked data with the stored data related to energy and memory.

\subsection{Organization of the paper}    
\label{SUBSEC:OUTLINE}
This paper is organized as follows: Section~\ref{SEC:RW} provides related works. Section~\ref{SEC:PA} outlines the architecture of our previous studies, while the detection mechanism of the novel study is presented in Section~\ref{SEC:DTM}. Sections~\ref{SEC:EXP5} present the results and evaluations. Finally, Section~\ref{ConFU} concludes with some final remarks and suggestions for future work.

\subsection{Motivation and Contribution}    
\label{SUBSEC:MOTIVATION}
IoT represents an interconnected network that links smart objects to the Internet. While a multitude of IoT devices are connected to the Internet, a significant portion of them lack robust security measures, rendering them vulnerable to various cybersecurity threats. These devices are often resource-constrained, making the application of conventional security protocols challenging. To address this, many researchers have proposed the implementation of intrusion detection systems at IoT gateways. However, the vulnerability of IoT devices persists if these gateway-based security mechanisms prove insufficient.

Therefore, there is a growing need for innovative solutions that can effectively safeguard IoT devices against resource-constrained attacks, particularly those aimed at depleting energy and memory resources. To enhance the security of IoT systems and services and defend against cyber threats that exploit the resource constraints of smart devices, an additional layer of protection specifically designed for resource-constrained IoT devices and networks is imperative.
Our primary contribution lies in conducting a comprehensive statistical analysis of the algorithms we've introduced in our previous studies~\cite{BICT2023,Fabulous2023,INISCOM2023}. This analysis serves to determine the most effective algorithms for detecting resource-constraint attacks. We achieve this by calculating various critical variables, including the probability of detection, probability of misdetection, probability of false alarm, and overall accuracy. This in-depth statistical examination allows us to provide valuable insights into the performance and effectiveness of these algorithms in defending against resource-constraint attacks.
Furthermore, we introduce a novel method for detecting resource constraint attacks. This innovative approach involves the analysis of captured packets at the protocol level, segmenting them into two categories: normal packets and attacked packets. Subsequently, ML techniques are employed to classify the nature of the attacked packet on the protocol, whether it involves an energy or memory-related attack. The classification process relies on comparing the extracted information from the attacked packets with our existing energy and memory attack pattern datasets. Furthermore, we concurrently monitor energy and memory metrics during the attack detection phase to determine the presence of an energy attack, memory attack, or both on the targeted smart devices. The calculation of accuracy and detection attack rate is also involved.

\section{Related Work}
\label{SEC:RW}
Several researchers have proposed methods to identify attacks on energy consumption. Valentina \emph{et al.}~\cite{Valentina2017} emphasized the significance of energy efficiency, promoting the use of smarter devices for sustainability. Efficient home automation control is crucial for minimizing energy losses, optimizing consumption based on specific needs, and ensuring effective system operation~\cite{FORD2017}. This investigation~\cite{kumar2021secure} evaluated established home energy management systems, highlighting differences in functionality and quality to identify opportunities for energy conservation, considering both behavioural and operational aspects. It's noteworthy that factors such as comfort, convenience, and security often influence the adoption of energy-efficient scenarios.
In another approach proposed by Shi \emph{et al.}, ~\cite{Shi2019}, a detection framework for IoT systems relies on energy consumption analysis. This method involves scrutinizing smart devices' energy usage, and categorising monitored devices' attack status, encompassing both cyber and physical attacks. The proposed two-stage process involves a short time window for initial attack detection and a longer time window for more refined attack identification.
Similarly, various authors have addressed the issue of detecting memory attacks. For instance, Mosli \emph{et al.}\cite{mosli2016automated} introduced a technique to detect malware by extracting three key features from memory images: Application Programming Interface (API) calls, registry activities, and imported libraries. These experiments were carried out on each feature separately, and the highest accuracy, reaching $96$\%, was achieved using the SVM classifier with the registry activities feature. In a subsequent work\cite{mosli2017behavior}, the same authors utilized process handles available in memory to detect malware. Their research found that malware typically employs process handles, mutants, and section handles. However, their approach achieved a modest accuracy when employing the RF classifier, slightly exceeding $91$\%.
Duan \emph{et al.}~\cite{duan2015detective} also introduced an approach for extracting live Dynamic-Link Library (DLL) features from memory, which was used to detect malware variants that employed the same DLLs. Their experiments resulted in an accuracy of $90$\%, achieved through the hidden naïve Bayes classifier.
Furthermore, Dai \emph{et al.}\cite{dai2018malware} proposed a malware detection and classification approach based on extracting memory images and converting them into fixed-size greyscale images. Features were extracted from these images using a gradient histogram, and these features were used for malware classification. An accuracy of $95.2$\% was achieved using a neural network classifier. In a previous work by the same authors, API calls from behaviour analysis and memory analysis were combined into a single vector to represent each sample. They employed a dataset consisting of $1200$ malware and $400$ benign files to train the SVM classifier. The research demonstrated that memory analysis could mitigate the limitations of behaviour analysis\cite{sihwail2019malware}. Indeed, the application of TinyML is indispensable in ensuring the security of smart devices against resource-constrained attacks. TinyML, which has effectively expanded the domain of ML to encompass low-power microcontrollers~\cite{tsoukas2022enhancing}, is a vital component in addressing the evolving landscape of IoT. As IoT devices become increasingly sophisticated in their analytical capabilities, networking, security, and decision-making challenges become more complex.

The existing constraints, including power, memory, and computational capacity limitations, are significantly impeding the creation of robust connections, the implementation of stringent security measures, and the fine-tuning of systems. In this context, TinyML necessitates the development of models capable of autonomous operation on edge nodes, ensuring both low-latency response and resilience in the face of these challenges.
While the studies presented in this paper have made substantial strides in resource-constraint attack detection for smart devices, the critical task of selecting the most effective detection mechanism remains. This study presents a comprehensive statistical analysis to address the limitations inherent in existing works.
In addition to algorithmic comparisons, we further contribute to the field by introducing a novel detection mechanism designed to identify attacks on smart devices. This mechanism involves real-time monitoring of packets, categorizing them into normal and attacked packets. Subsequently, ML algorithms are applied exclusively to the attacked packets, allowing us to precisely specify whether the attack is related to energy consumption or memory usage.
Through this innovative approach, we aim to enhance the overall security posture of smart devices. The comprehensive statistical analysis not only evaluates existing algorithms but also introduces a novel mechanism to bolster the detection capabilities, providing a more holistic understanding of the effectiveness of these techniques against resource-constraint attacks on smart devices.

\begin{figure}[!h]
	\includegraphics[width=\linewidth]{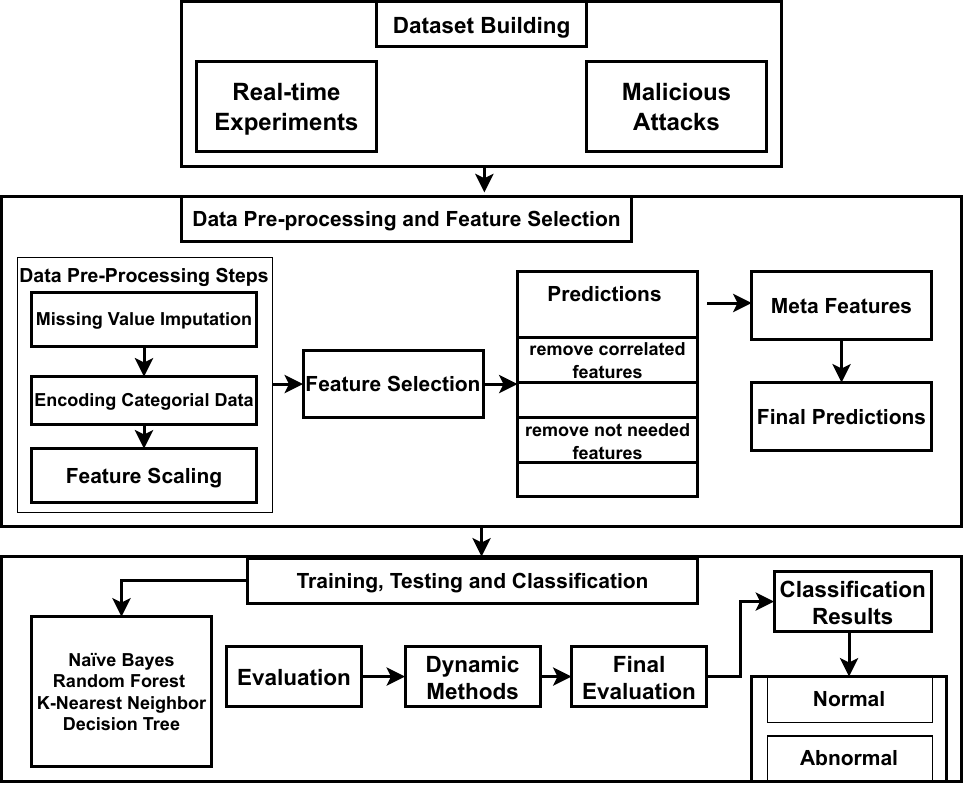}
	\centering
	\caption{Proposed Architecture.} 
	\label{FIG:dataset}
\end{figure}
\section{Proposed Architecture and Dataset selections}
\label{SEC:PA}

The proposed system architecture for both techniques used to detect resource constraint attacks is described as follows and shown in Figure~\ref{FIG:dataset}:

The first technique encompasses a monitoring system to observe the energy consumption and memory usage of smart devices in various system states, such as \emph{Idle}, \emph{Normal}, and \emph{Abnormal}. Subsequently, the detection mechanism compares the normal energy consumption and memory usage behaviour with abnormal behaviour to identify potential attacks.
The second algorithm primarily employs the TinyML technique to distinguish between normal and abnormal behaviour in the context of resource constraints within smart devices. This system follows four key phases: dataset creation, data pre-processing, feature selection, and resource constraint attack detection. During dataset creation, real-time experiments are conducted to collect data on energy consumption and memory usage in the presence and absence of attacks. Features are extracted from the recorded energy consumption, memory usage readings, and simulated attack signals. These features, representing all samples, are included in a dataset for further pre-processing. The second phase involves data pre-processing and feature selection, focusing on extracting energy and memory values from the dataset. TinyML techniques, such as Naive Bayes (NB), Decision Tree (DT), K Nearest Neighbor (K-NN), and RF, are employed for dataset training.

A classifier is utilized to analyze normal behaviour in these devices to detect resource constraint attacks on smart devices. It then compares this behaviour with abnormal patterns, enabling the detection of anomalies. Incoming signals are subjected to a classification process to determine their authenticity or whether they may be considered potentially spoofed. The probabilities associated with these signals are evaluated during the protection phase.

For the dataset and data pre-processing. This study leverages a dataset developed in our prior work~\cite{alwaisi2023optimized}. Real-time experiments were conducted on authentic smart devices to gauge resource metrics such as energy consumption and memory usage, facilitating the creation of a dataset representing normal and abnormal behaviours. Energy consumption was measured using a smart circuit~\cite{BICT2023}, while memory usage was monitored through various Python and C language libraries~\cite{INISCOM2023}. Abnormal behaviour was induced for dataset collection, including Distributed Denial-of-Service (DDoS) and Energy Consumption-Distributed Denial-of-Service (EC-DDoS) attacks. Extracted features from the energy consumption dataset include device information, ports, protocols, and energy consumption values. Similarly, features extracted from memory monitoring encompass device details, ports, protocols, and memory usage metrics. For pre-processing the dataset the dataset underwent meticulous pre-processing, entailing the identification and elimination of null, unknown, and noisy values during the missing value identification step.

\section{Resource constraints Detection Method based on Machine Learning}
\label{SEC:DTM}

This section delves deeper into our ML approach, a pivotal element in our paper. Our objective is to detect attacks on the protocols of smart devices effectively. To achieve this, we meticulously categorise protocol data into two distinct classes: normal and abnormal. The abnormal data is then subjected to a thorough analysis, with a primary focus on resource usage, particularly memory and energy consumption. The aim here is to precisely identify the nature of the attack, whether it involves energy, memory, or falls into another category of malicious activity. In this context, the choice of ML algorithms plays a pivotal role in the effectiveness of our approach. We have opted for two highly regarded algorithms: RF and SVM~\cite{hussain2020machine,rana2020performance}. Both have proven their worth in various domains, and their application offers several significant advantages.

To achieve this, we employ a packet capture tool to perform a comparative analysis between the captured attack packets and the regular data packets. This process involves identifying patterns and rules within the attack data and translating them into distinctive features. Through the analysis of the protocol data, we sent malicious attacks that affect the energy consumption and memory usage of the smart devices, e.g., DDoS attacks. Then, we collect the normal and abnormal packet data for further analysis. These extracted data are then utilized as inputs for training the ML model to specify whether there is an energy or memory attack on the smart devices.

\begin{figure}[!h]
	\includegraphics[width=\linewidth]{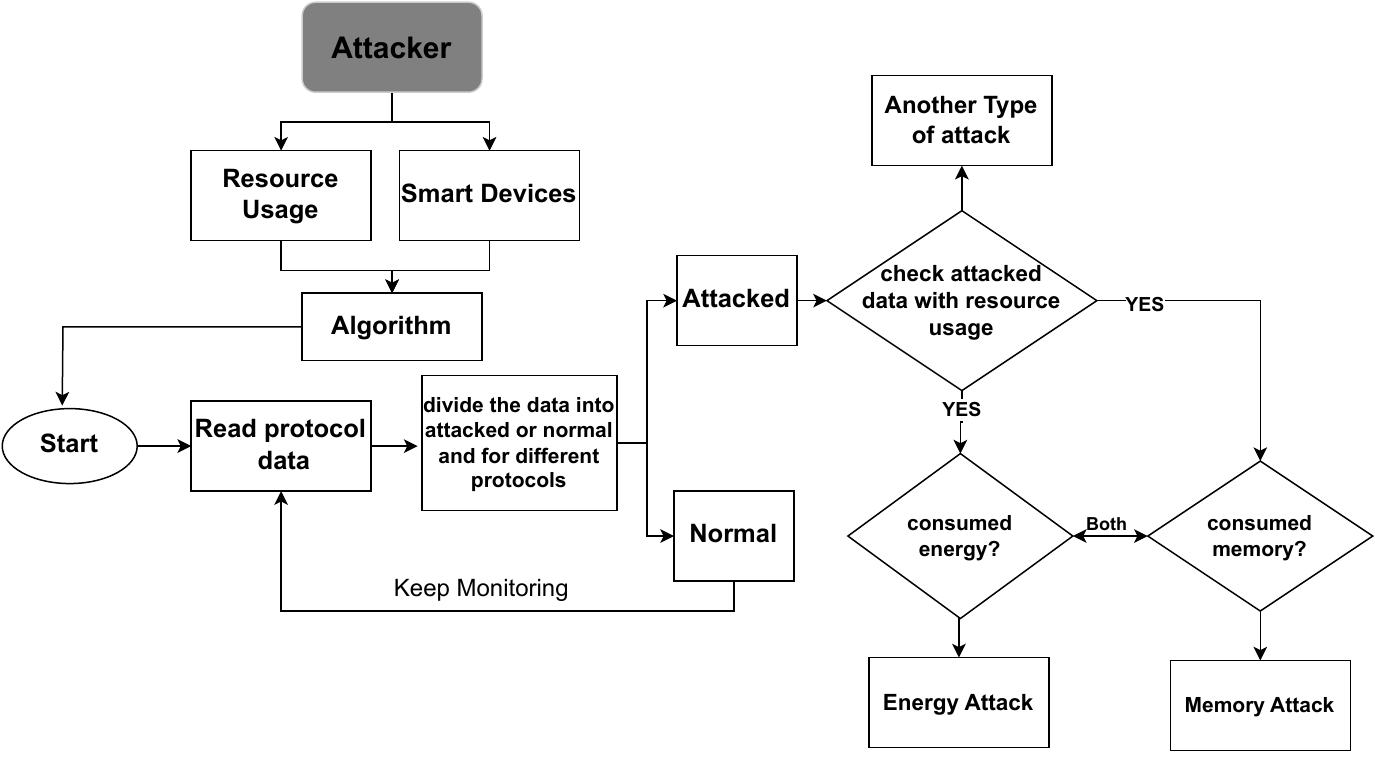}
	\centering
	\caption{Testing Environment and a technique to detect resource usage attacks in smart devices.} 
 
	\label{FIG:algorithm}
\end{figure}
The packet capture tool takes advantage of tcpdump, and Wireshark, a robust network data analysis tool well-regarded among internet system administrators. Tcpdump offers extensive flexibility, allowing users to create custom filters for packet analysis. For example, using a straightforward command like (tcpdump -i en1 -v tcp), users can capture Transmission Control Protocol (TCP) protocol network packets from the \emph{en1} network interface.
Once it accumulates sets of attack packets related to resource constraint attacks, the tool initiates a comparative analysis by juxtaposing these attack packets against regular traffic for TCP and User Datagram Protocol (UDP). This analytical process reveals distinct characteristics associated with each attack mode. Notably, variations emerge between attacked data packets and normal data packets across different protocols. For instance, in the case of the TCP protocol, differences manifest in the packet sequence—regular for normal cases and random during attacks. Similarly, IP sources exhibit regular patterns for normal scenarios, with multiple destination IPs, while abnormal cases present confusion and a single destination IP. Furthermore, Packet Identification in normal cases features varying bits, whereas in attacked cases, the identification aligns with packets sent by the attacker.
Contrastingly, in the UDP protocol, deviations are observed in the port number—normal cases display a specific port number, while abnormal behaviour involves random ports. Additionally, packet length demonstrates irregularities in normal scenarios, whereas abnormal cases exhibit consistent lengths.

In summary, based on the comparative analysis of regular protocol data and protocol attack data, the general characteristics of protocols can be summarized as follows:

\begin{equation}
    PROT_{protocol}= (NUM_{packet}, LEN_{packet}, IDEN_{packet}, PORT)
\end{equation}

For the TCP protocol, the characteristics will be as follows: 

\begin{equation}
  TCP_{protocol}  = (TCP_{NUM}, TCP_{LEN}, TCP_{IDEN})
\end{equation}

For the UDP, the characteristics will be defined as follows: 

\begin{equation}
  UDP_{protocol}  = (UDP_{NUM}, UDP_{LEN}, UDP_{port})
\end{equation}

After dividing the protocol packet into normal and attacked data packets, we then use ML on the attacked packets to specify whether there is an energy or memory attack on the smart devices by comparing the fetched data with the energy and memory of the smart devices. 
Thus, RF represents a crucial ensemble learning technique built upon the Bagging framework, primarily designed for addressing classification and regression challenges. Decision trees serve as the base model for the Bagging process. RF offers notable advantages, including straightforward parallelization and enhanced prediction accuracy, all achieved without substantially increasing computational overhead. In contrast, SVM is a versatile classification and regression tool known for its proficiency with linear and non-linear data. It excels in high-dimensional spaces, a trait particularly valuable in our analysis of intricate protocol data. SVM's ability to determine optimal hyperplanes for distinguishing between different classes provides an effective means of identifying abnormal data patterns.
\[
\text{Decision} = \begin{cases} 
\text{Attacked} & \text{if } \sum_{i=1}^{n} w_i x_i + b > \text{Threshold} \\
\text{Normal} & \text{otherwise}
\end{cases}
\]
$w_i$ represents the weights assigned to each feature ($x_i$) in the feature vector, $b$ is the bias term, $n$ is the number of features, and the threshold is a decision boundary separating attacks from normal packets. Therefore, the normal packets and the feature ($X_i$) are learned during the training phase of the ML model. The features ($x_i$) include various characteristics of the packet data, such as packet length, source/destination IP addresses, protocol type, etc.  

The combined use of RF and SVM in our training and evaluation processes enhances the reliability of our results. These algorithms work in synergy to handle the complexities of resource-constrained attacks. RF's adaptability and SVM's proficiency in classifying data types play a vital role in our approach.
Therefore, based on the categorization of attack protocols, the attack detection models are categorized into three distinct groups: the TCP attack detection model, the UDP attack detection model, and a general attack detection model. The specific training model steps are outlined as follows:

\begin{enumerate}
    \item Execute feature extraction, data format conversion, and dimensional reconstruction on the acquired attack data to create efficient datasets, retaining the relevant feature values.
    \item Partition the training dataset into K subsets of equal size, using K-1 subsets for model training and the one designated for cross-validation.
    \item Reiterate the model using various K values and determine the number of decision trees compared to the highest average accuracy among the different K values as the number of decision trees in the RF algorithm.
\end{enumerate}

\subsection{Data Collection Process}

In our pursuit of a comprehensive security solution, we extend our methodology by not only evaluating algorithms but also by conducting an in-depth analysis of the collected data. The data collection process involves capturing information from the protocols commonly used by smart devices, including TCP and UDP. This approach allows us to curate datasets that serve as the foundation for both our statistical analysis and ML model training.
Our system distinguishes between normal and abnormal behaviour through the collected data. In the abnormal behaviour scenario, we intentionally expose smart devices to various malicious attacks, simultaneously collecting data. This process facilitates a robust evaluation of the system's response to attacks under different conditions. Subsequently, the attacked data is meticulously compared with the energy and memory usage of the smart devices. This comparative analysis aids in precisely specifying whether the attack pertains to energy consumption or memory usage, offering insights into the nature of the threat.
For data collection, we employ specialized tools, such as tcpdump, to capture packets from TCP and UDP protocols. Tcpdump operates from the command line, providing the capability to capture and display detailed packet information, making it suitable for scripting and automation. Additionally, we utilize Wireshark, a widely-used network protocol analyzer with a graphical user interface, to monitor and capture further information, enhancing our understanding of the network traffic and facilitating more nuanced analysis.
By leveraging real-world scenarios and intentional exposure to attacks, our approach ensures a holistic understanding of smart device security, paving the way for effective defences against resource-constrained attacks.

\begin{enumerate}
    \item Dataset preparation: During this phase, the dataset has been acquired. The initial step involves gathering data on normal activities, ensuring their integrity without any associated attacks. This phase encompasses subphases focusing on IoT sensor devices, packet analysis, and subsequent data.  

    \item Data collection: At this stage, the data for the different protocols are collected for normal and abnormal behaviours. The analysis of attacks took place at various time intervals every 2, 3, 5, and 10 seconds, respectively. The data collection involved gathering information both before and after the occurrence of the two specified types of attacks. This data was meticulously formatted and structured in CSV format to calculate further information. Subsequently, the data pre-processing phase commences, involving the capture and analysis of network packets using Wireshark. This process aims to extract attributes such as timestamp, source IP address, destination IP address, source port, destination port, protocol, and bytes. Additionally, IP addresses are converted into numerical values.
 
    \item Data cleaning and parsing: In this phase, we grapple with a substantial volume of data, necessitating thorough cleaning and extraction of distinct features from the dataset. We remove duplicate entries and eliminate superfluous information to ensure data quality. Subsequently, the refined data is stored in the database, poised for further computational analysis. 
$\text{Clean}(D) \rightarrow \{D_i \mid D_i \in D, i = 1, 2, \ldots, N\}$ this indicates that the cleaned data is a set of elements $D_i$ 
  for each index $i$ from $1 to N$ , where each $D_i$ is selected from the original array $D$ and contains non-redundant data.
\begin{equation}
D' = \text{Clean}(D)
\end{equation}
where $D'$ is the cleaned array.

   \item Feature extraction: the information collected for this experiment encompasses diverse data attributes, including packet sequence, IP sources, packet identification, port numbers, and packet length. These data play a pivotal role in the initial stages of distinguishing between normal and abnormal behaviours. Additionally, they provide essential insights into energy and memory usage patterns. 
\end{enumerate}

\section{Experimentation and Discussion}
\label{SEC:EXP5}
\subsection{Test Results and Discussions}
\label{TestResult}

Determining whether the detection data corresponds to a resource constraints attack constitutes a classification problem. To assess the experimental outcomes, we utilize evaluation criteria such as the false positive rate, true positive rate, false detection rate, and accuracy.

The accuracy (ACC) calculation pertains to the ratio of correct predictions (Attack traffic and Normal traffic) to the total number of observations, and it is computed as follows:
\begin{equation}
    ACC=\frac{TP+TN}{N}
\end{equation}

Moreover, it is important to calculate some important variables. For example, the false positive rate (FPR) denotes the percentage of normal behaviour inaccurately recognized as attack data, and it is defined as:

\begin{equation}
    FPR=\frac{FP}{(FP+TN)}
\end{equation}

The true positive rate (TPR), also known as sensitivity or recall, is the proportion of true positives to the total number of observations predicted as positive.

\begin{equation}
    TPR=\frac{TP}{TP+FN}
\end{equation}

The false detection rate (FDR) is defined as ($1- ACC$), and it is defined as follows:

\begin{equation}
  FDR=\frac{FP + FN}{N}  
\end{equation}

TP (True Positive) denotes instances where the model correctly predicts attack traffic as an attack.
TN (True Negative) indicates instances where the model correctly predicts normal traffic as normal.
FP (False Positive) occurs when the model incorrectly predicts normal traffic as an attack.
FN (False Negative) represents instances where the model incorrectly predicts attack traffic as normal. Finally, N is the number of samples.
\begin{figure}[!h]
	\includegraphics[width=\linewidth]{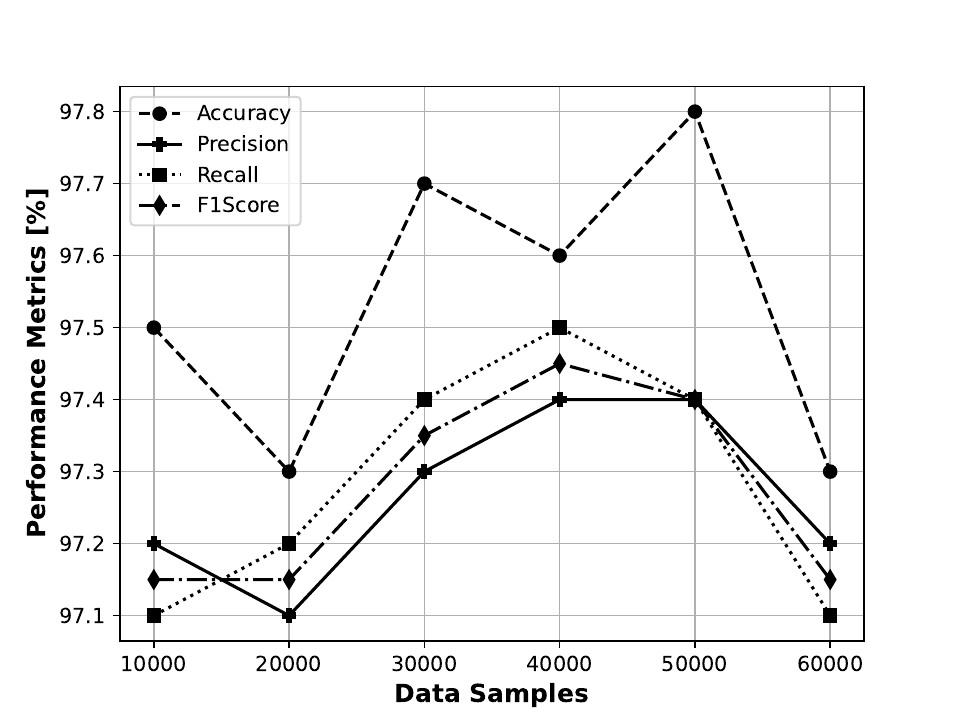}
	\centering
	\caption{Accuracy, Precision, Recall, F1Score of the proposed models. }

	\label{FIG:ev}
\end{figure}
After the RF model is trained using the training dataset, we proceed to create the test dataset by integrating the remaining attack data packets with normal traffic. This involves systematically combining normal and attack traffic samples, determining the classification of each sample, and carefully controlling the flow to maintain the desired ratio of normal to attack traffic. Simultaneously, we utilize the LIBSVM library for processing data using the SVM algorithm, and these outcomes are subsequently compared with the results obtained from the RF model. Therefore, after detecting if there is an attack across the mentioned protocol, we then take the attacked data to specify the main reason behind such a type of attack; we will compare the fetched attacked data from a protocol and then compare its values with energy or memory measurements to specify if the attack is on memory or energy or effect both energy consumption and memory usage of the smart devices. Table~\ref{table:firsttable} presents the findings for detecting if there is an attack across the different mentioned protocols:

\label{table:firsttable}

\begin{table}
\renewcommand{\arraystretch}{1.8}
\centering
\footnotesize

\caption{Comparison of Detection Performance Metrics for SVM and RF Algorithms across TCP and UDP Protocols.  }

\label{table:firsttable}
\begin{tabularx}{\textwidth}{|>{\centering\arraybackslash}l|>{\centering\arraybackslash}l|>{\centering\arraybackslash}X|>{\centering\arraybackslash}X|>{\centering\arraybackslash}X|>{\centering\arraybackslash}X|}
\hline
\textbf{Algorithm} & \textbf{Protocol} & \textbf{ACC [\%]} & \textbf{TPR [\%]} & \textbf{FPR [\%]} & \textbf{FDR [\%]} \\ \hline
\multirow{2}{*}{\textbf{SVM}} & TCP & 94.5  & 93.75 & 5.70 & 5.5  \\ \cline{2-6} 
                              & UDP & 98.8  & 95.7  & 0.04 & 0.05 \\ \hline
\multirow{2}{*}{\textbf{RF}}  & TCP & 99.50 & 95.83 & 0    & 0.50 \\ \cline{2-6} 
                              & UDP & 99.12 & 91.17 & 0    & 0.01 \\ \hline
\end{tabularx}
\end{table}

The RF algorithm consistently exhibits higher accuracy across all protocol types compared to the SVM technique, as illustrated in Table~\ref{table:firsttable}. Specifically, RF achieves an accuracy of approximately $99.50$\% for TCP and $99.12$\% for UDP. In contrast, SVM yields an accuracy of $94.5$\% for TCP and $98.8$\% for UDP.
Examining false positive rates, the SVM algorithm registers $5.70$ for TCP and $0.04$\% for UDP. Conversely, the RF algorithm demonstrates lower false positive rates, recording values of $0$\% for both protocols. Simultaneously, the False Discovery Rate (FDR) for RF is $0.50$\% for TCP and $0.01$\% for UDP. In comparison, the SVM algorithm achieves an FDR of $5.5$\% for TCP and $0.05$\% for UDP.
Finally, considering the True Positive Rate (TPR), RF attains $95.83$\% for TCP and $91.17$\% for UDP, while SVM records TPR values of $93.75$\% for TCP and $95.7$\% for UDP. 
 
Furthermore, Figure~\ref{FIG:ev} depicts a comprehensive analysis of performance metrics, encompassing accuracy, recall (formulated as $TP / (TP + FN)$), precision (expressed as $TP / (TP + FP)$), and F1 score (calculated as $2 \times \frac{{\text{{Precision}} \times \text{{Recall}}}}{{\text{{Precision}} + \text{{Recall}}}}$). This analysis spans multiple models evaluated across varying data sample sizes, specifically $10,000$, $20,000$, $30,000$, $40,000$, $50,000$, and $60,000$ instances.

\subsection{Statistical Analysis Results}
\label{SEC:RS}

We conducted our simulation on an Intel Core i7-10750H CPU, operating at 2.60 GHz, with 16.0 GB of memory.

We employed four evaluation metrics to appraise the effectiveness of the proposed model. These metrics include the Probability of Detection ($P_d$), Probability of False Alarm ($P_{fa}$), Probability of Misdetection ($P_{md}$), and Accuracy ($ACC$). The calculations for these metrics were carried out using the following formulas:

 \begin{equation}
        P_d=\frac{T_P}{T_P+F_N}
    \end{equation}

 \begin{equation}
        P_{fa}=\frac{F_P}{T_F+F_N}
    \end{equation}

 \begin{equation}
        P_{md}=\frac{F_N}{T_N+F_P}
    \end{equation}

 \begin{equation}
        ACC=\frac{T_P + T_N}{T_P+ T_N+F_P+F_N}
    \end{equation}

In the context of our evaluation, we used the following notation: TP represents the count of correctly predicted malicious flows, TN corresponds to the count of accurately predicted normal flows, FP stands for the count of erroneously predicted malicious flows, and FN represents the count of improperly predicted normal flows.
To assess the performance of the proposed dynamic methods, we conducted a simulation analysis and compared the results against those of the different devices chosen and the different techniques used in our previous studies. 

\begin{figure}[!h]
	\includegraphics[width=\linewidth]{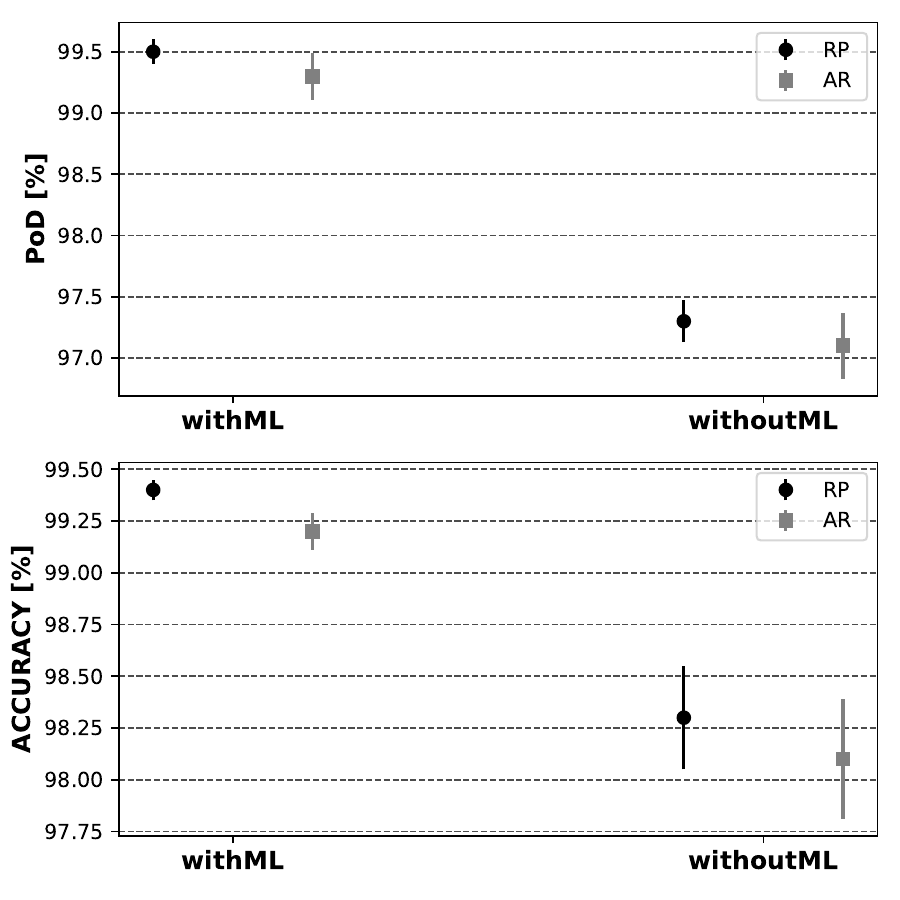}
	\centering
	\caption{Evaluation Results in terms of Accuracy and Probability of Detection (PoD).}
	\label{FIG:poDACC}
\end{figure}

\begin{figure}[!h]
	\includegraphics[width=\linewidth]{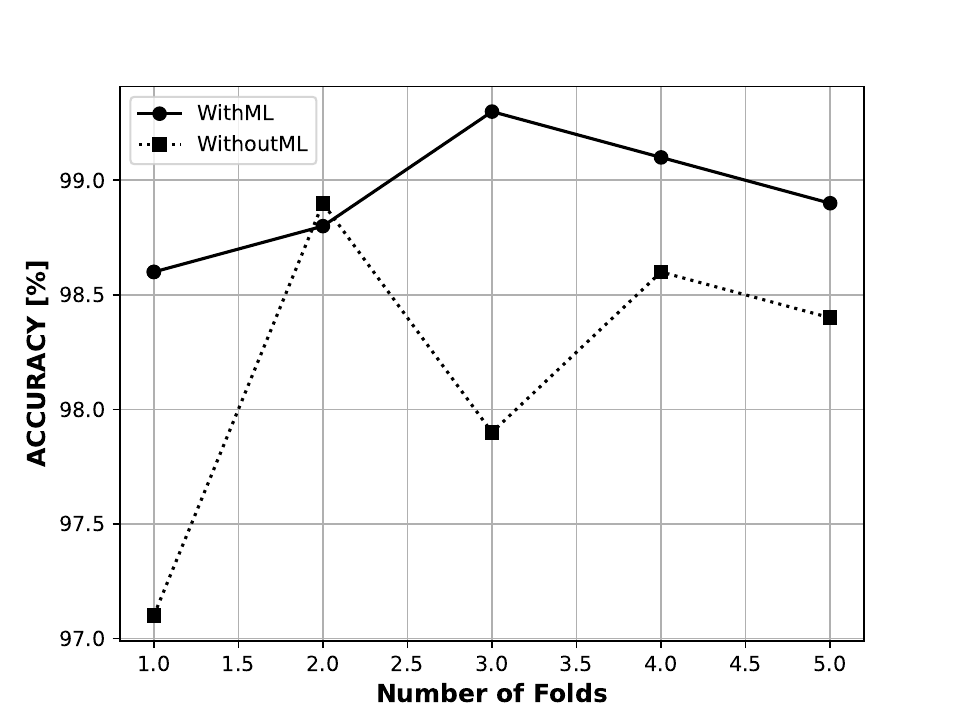}
	\centering
	\caption{The accuracy for various numbers of folds. }

	\label{FIG:kfold}
\end{figure}

Figure~\ref{FIG:poDACC} presents the results of the proposed methods and two different algorithms applied to various smart devices, focusing on accuracy. Notably, the proposed algorithm for detecting resource-constrained attacks using TinyML in Raspberry Pi (RP) and Arduino (AR) devices showcases superior accuracy compared to the other algorithms. As evident from the data, the resource-constrained attack detection using RP and AR devices with Tiny ML achieved an impressive accuracy of $99.4$\% and $99.2$\%, respectively. In contrast, the algorithms for detecting resource-constrained attacks without the utilization of Tiny ML yielded accuracies of $98.3$\% for RP and $98.1$\% for Arduino. The latter represents the lowest accuracy among the considered algorithms. Moreover, we employed k-fold cross-validation to provide a more detailed analysis of the algorithmic performance. Given the constraints of small datasets related to smart devices, we employed 5-fold cross-validation independently for each algorithm, a total of $\approx 66000$ samples divided between these five folds. This approach helps robustly assess our algorithms' accuracy across different subsets of the data, as shown in Figure~\ref{FIG:kfold}. 

The accuracy across the folds was consistently good, with values of $98.6$\% $98.3$\%, $99.3$\%, $99.1$\%, and $98.9$\%, respectively, for the algorithm that uses TinyML, and $97.1$\% $97.9$\%, $98.8$\%, $98.6$\%, and $98.4$\% for the other algorithm, respectively.
This enhanced accuracy contributes to the reliability and generalizability of our models, which is particularly crucial when dealing with limited data on smart devices.

Figure~\ref{FIG:poDACC} also presents the results of the proposed methods and two different algorithms applied to various smart devices, focusing on the probability of detection.
As can be seen, the proposed algorithm for detecting resource-constrained attacks using TinyML in RP and AR in terms of probability of detection has a slight difference. TinyML has the
highest probability of detection of $99.4$\% for the RP and $99.2$\% for the AR. The other algorithm has a probability of detection of $97.3$\% for the RP and $97.1$\% for the AR. 

\begin{figure}[!h]
	\includegraphics[width=\linewidth]{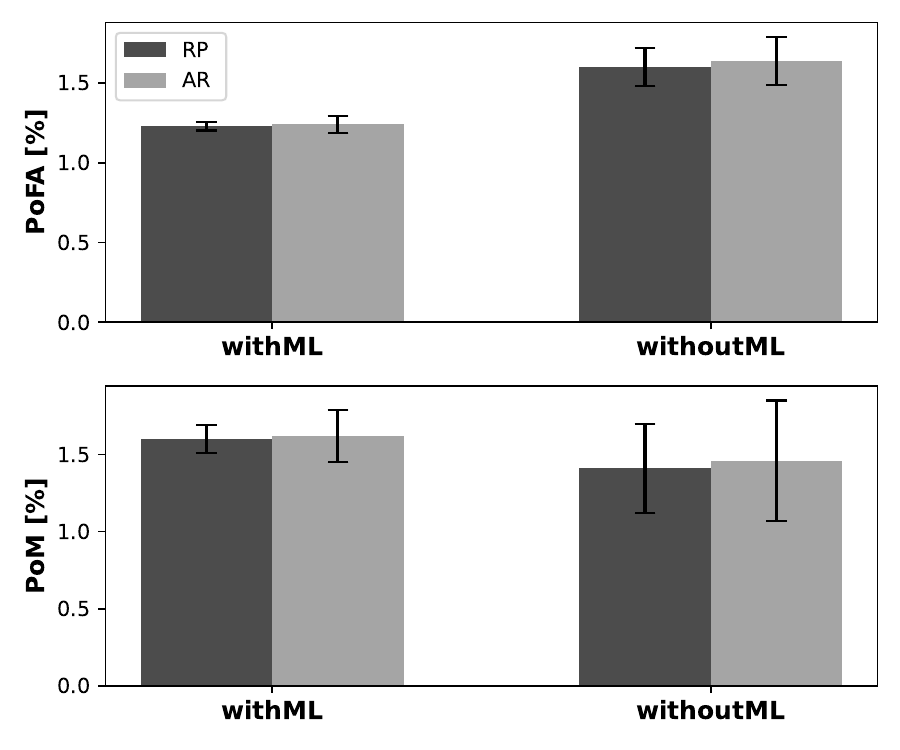}
	\centering
	\caption{Evaluation Results in terms of PoF and PoM. }

	\label{FIG:PoFPoM}
\end{figure}

Figure~\ref{FIG:PoFPoM} depicts the outcomes of PoFA and PoM for detecting resource constraint attacks. We utilized TinyML for the RP and AR models, while the other approach did not employ TinyML techniques. Specifically, this analysis examines the probability of misdetection (PoM).
As one can observe, the PoM of the RP and AR using TinyML have an acceptable probability of misdetection; however, the lowest
probability of misdetection belongs to the other algorithm. The proposed methods have a probability of misdetection of $1.60$\% for both smart devices by using TinyML to detect resource constraint attacks, while the other algorithm for both device models has a probability of misdetection of $1.46$\%. We also examine these results in terms of the probability of false alarms. It is clear that RP and AR, utilizing TinyML, achieve superior performance in terms of the probability of false alarms compared to the alternative algorithm, which does not employ TinyML.
To be precise, methods used TinyML yield a probability of false alarm of $1.23$\%. For RP and AR, specifically, these values stand at $1.2$\% and $1.23$\%, respectively. In contrast, the other algorithms, when not utilizing TinyML, yield higher probabilities of false alarms for AR and RP, which are $1.6$\% and $1.64$\%, respectively, as shown in Table~\ref{tab:evaluation}.
\begin{figure}[!h]
	\includegraphics[width=\linewidth]{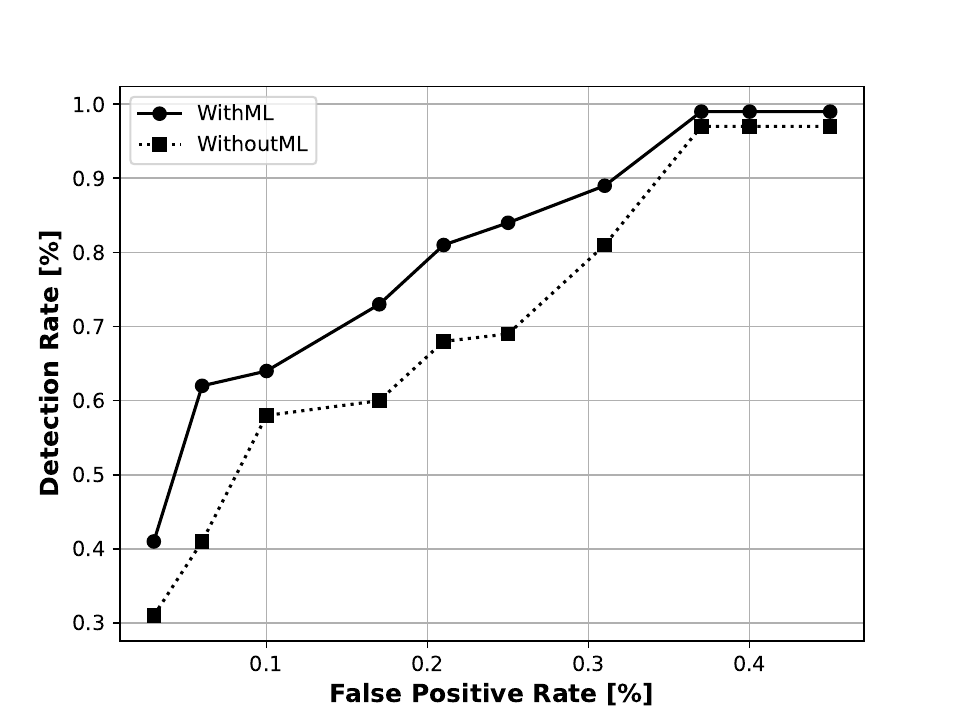}
	\centering
	\caption{Comparative Analysis of False Positive Rate and Detection Rate. }

	\label{FIG:fpdt}
\end{figure}
The count of false positive predictions ($f_p$) is a crucial metric in algorithm evaluation, as depicted in Table~\ref{tab:evaluation}. This metric compares the instances where the algorithm wrongly predicts a positive outcome to the total number of samples predicted as negative. In our proposed algorithm utilizing TinyML, the false positive rate is $10.9$ per second. In contrast, the alternative algorithm records a higher rate of $16$ false positives per second. Moreover, achieving a balance between the detection and false positive rates necessitates a trade-off. With an increase in the false positive rate, there is a corresponding rise in the detection rate. Hence, the false positive rate becomes a key metric for evaluating the system's effectiveness. As presented in Figure~\ref{FIG:fpdt} the provided information includes the detection rate corresponding to various false positive rates for both the algorithms, namely withML and withoutML. The presented results show different instances when resource-constraint attacks flooded smart devices. The threshold for normal cases was shown in different instances to show the effect of the threshold on the final results of FPR and DR. The first algorithm outperforms the second one. The results reveal two discernible trends: the detection rate is influenced by the false positive rate, and the proposed system consistently exhibits notable improvements in the detection rate.

\begin{table}[htbp]
\renewcommand{\arraystretch}{2}
	\centering
    \setlength{\leftmargini}{10pt}
    \begin{footnotesize}
	\caption{Evaluation Results of the Proposed Detection Algorithms.}

 \label{tab:evaluation}
\begin{tabular}{|c|c|c|c|c|c|c|ll}
\cline{1-7}
\textbf{Methods} & \textbf{Devices} & \textbf{ACC(\%)} & \textbf{PoD(\%)} & \textbf{PoMD(\%)} & \textbf{PoFA(\%)} & \textbf{FP(s)} &  &  \\ \cline{1-7}
\multirow{2}{*}{\textbf{With TinyML}}    & RP & 99.4 & 99.5 & 1.60 & 1.23 & \multirow{2}{*}{10.9} &  &  \\ \cline{2-6}
                                         & AR & 99.2 & 99.3 & 1.64 & 1.24 & &  \\ \cline{1-7}
\multirow{2}{*}{\textbf{Without TinyML}} & RP & 98.3 & 97.3 & 1.46  & 1.60  &  \multirow{2}{*}{16} &  &  \\ \cline{2-6}
                                         & AR & 98.1 & 97.41& 1.46 & 1.64 &  &  \\ \cline{1-7}
                                        
\end{tabular}
\end{footnotesize}
\end{table}

\section{Conclusion and Future work}
\label{ConFU}

In recent years, the escalating interest in the IoT and the security of its smart devices has given rise to significant technological advancements. These innovations have paved the way for many techniques and methods designed to identify and mitigate vulnerabilities within IoT systems. Yet, despite the progress, this field continues to grapple with formidable challenges and limitations, notably marked by the persistence of high rates of misdetection and false alarms.
Interest in detecting resource-constraint attacks in IoT systems has significantly increased in the last decade, leading to notable technological advancements. Several techniques and methods have been proposed to identify and address these vulnerabilities. However, this field of study still grapples with certain challenges and limitations, including high misdetection rates and false alarms. This work addresses this gap by proposing a novel detection mechanism for resource-constraint attacks in smart devices.
Our approach involves calculating various variables on the protocol, such as fetching the protocol data and dividing them into normal or attacked packets. We then apply ML techniques to specify if it's an energy or memory attack by comparing the attacked data with real measurements of energy and memory. The final results demonstrate a very high accuracy using RF and SVM. Specifically, the accuracy for RF exceeds $99.7$\%, while for SVM, it surpasses $98.8$\%. These results hold true across different protocols.

As a comparative reference for our previous works and detection mechanisms in our previous algorithms, the first algorithm employed for detecting resource-constraint attacks with TinyML achieves an accuracy of $99.4$\% and $99.2$\%, a probability of detection of $99.5$\% and $99.3$\%, a probability of misdetection of $1.62$\%, and a probability of false alarm of $1.23$\%. Conversely, the other algorithm, which does not utilize TinyML, exhibits an accuracy of $98.3$\% and $98.1$\%, a probability of detection of approximately $97.4$\%, a probability of misdetection of about $1.46$\%, and a probability of false alarm of approximately $1.64$\%. These findings underscore the effectiveness of our proposed detection mechanism in addressing the challenges associated with resource-constraint attacks in IoT systems. In future work, we will delve deeper into utilizing ML techniques to detect other types of attacks that affect the resources of smart devices.

\bibliographystyle{splncs04}
\bibliography{bibliography}

\begin{thebibliography}{10}
\providecommand{\url}[1]{\texttt{#1}}
\providecommand{\urlprefix}{URL }
\providecommand{\doi}[1]{https://doi.org/#1}

\bibitem{Fabulous2023}
Al-Waisi, Z., Soderi, S., De~Nicola, R.: Detection of energy consumption cyber attacks on smart devices. vol.~12, p.~1927. EAI-SPRINGER (2023)

\bibitem{INISCOM2023}
Alwaisi, Z., Soderi, S., De~Nicola, R.: Mitigating and analysis of memory usage attack in ioe system. In: Vo, N.S., Tran, H.A. (eds.) Industrial Networks and Intelligent Systems. pp. 296--314. Springer Nature Switzerland, Cham (2023). \doi{$10.1007/978-3-031-47359-3_22$}

\bibitem{BICT2023}
Alwaisi, Z., Soderi, S., Nicola, R.D.: Energy cyber attacks to smart healthcare devices: A testbed. In: Chen, Y., Yao, D., Nakano, T. (eds.) Bio-inspired Information and Communications Technologies. pp. 246--265. Springer Nature Switzerland, Cham (2023). \doi{$10.1007/978-3-031-43135-7_24$}

\bibitem{alwaisi2023optimized}
AlWaisi, Z.A.: Optimized monitoring and detection of internet of things resources-constraints cyber attacks  (2023). \doi{https://e-theses.imtlucca.it/392/}

\bibitem{dai2018malware}
Dai, Y., Li, H., Qian, Y., Lu, X.: A malware classification method based on memory dump grayscale image. Digital Investigation  \textbf{27},  30--37 (2018)

\bibitem{duan2015detective}
Duan, Y., Fu, X., Luo, B., Wang, Z., Shi, J., Du, X.: Detective: Automatically identify and analyze malware processes in forensic scenarios via dlls. In: 2015 IEEE International Conference on Communications (ICC). pp. 5691--5696. IEEE (2015)

\bibitem{Valentina2017}
Fabi, V., Spigliantini, G., Corgnati, S.P.: Insights on smart home concept and occupants’ interaction with building controls. Energy Procedia  \textbf{111},  759--769 (2017). \doi{https://doi.org/10.1016/j.egypro.2017.03.238}, \url{https://www.sciencedirect.com/science/article/pii/S1876610217302680}, 8th International Conference on Sustainability in Energy and Buildings, SEB-16, 11-13 September 2016, Turin, Italy

\bibitem{FORD2017}
Ford, R., Pritoni, M., Sanguinetti, A., Karlin, B.: Categories and functionality of smart home technology for energy management. Building and Environment  \textbf{123},  543--554 (2017). \doi{https://doi.org/10.1016/j.buildenv.2017.07.020}

\bibitem{hussain2020machine}
Hussain, F., Hussain, R., Hassan, S.A., Hossain, E.: Machine learning in iot security: Current solutions and future challenges. IEEE Communications Surveys \& Tutorials  \textbf{22}(3),  1686--1721 (2020)

\bibitem{Kamaldeep2020}
Kamaldeep, Dutta, M., Granjal, J.: Towards a secure internet of things: A comprehensive study of second line defense mechanisms. IEEE Access  \textbf{8},  127272--127312 (2020)

\bibitem{Khor2021}
Khor, J.H., Sidorov, M., Woon, P.Y.: Public blockchains for resource-constrained iot devices—a state-of-the-art survey. IEEE Internet of Things Journal  \textbf{8}(15),  11960--11982 (2021). \doi{10.1109/JIOT.2021.3069120}

\bibitem{kumar2021secure}
Kumar, A., Sharma, S., Goyal, N., Singh, A., Cheng, X., Singh, P.: Secure and energy-efficient smart building architecture with emerging technology iot. Computer Communications  \textbf{176},  207--217 (2021)

\bibitem{Liu18}
Liu, X., Liu, Y., Liu, A., Yang, L.T.: Defending on–off attacks using light probing messages in smart sensors for industrial communication systems. IEEE Transactions on Industrial Informatics  \textbf{14}(9),  3801--3811 (2018). \doi{10.1109/TII.2018.2836150}

\bibitem{mosli2016automated}
Mosli, R., Li, R., Yuan, B., Pan, Y.: Automated malware detection using artifacts in forensic memory images. In: 2016 IEEE Symposium on Technologies for Homeland Security (HST). pp.~1--6. IEEE (2016)

\bibitem{mosli2017behavior}
Mosli, R., Li, R., Yuan, B., Pan, Y.: A behavior-based approach for malware detection. In: IFIP International Conference on Digital Forensics. pp. 187--201. Springer (2017)

\bibitem{pahl2018all}
Pahl, M.O., Aubet, F.X.: All eyes on you: Distributed multi-dimensional iot microservice anomaly detection. In: 2018 14th International Conference on Network and Service Management (CNSM). pp. 72--80. IEEE (2018)

\bibitem{Pahl18}
Pahl, M.O., Aubet, F.X., Liebald, S.: Graph-based iot microservice security. In: NOMS 2018 - 2018 IEEE/IFIP Network Operations and Management Symposium. pp.~1--3 (2018). \doi{10.1109/NOMS.2018.8406118}

\bibitem{pajouh2016two}
Pajouh, H.H., Javidan, R., Khayami, R., Dehghantanha, A., Choo, K.K.R.: A two-layer dimension reduction and two-tier classification model for anomaly-based intrusion detection in iot backbone networks. IEEE Transactions on Emerging Topics in Computing  \textbf{7}(2),  314--323 (2016)

\bibitem{rana2020performance}
Rana, V.K., Suryanarayana, T.M.V.: Performance evaluation of mle, rf and svm classification algorithms for watershed scale land use/land cover mapping using sentinel 2 bands. Remote Sensing Applications: Society and Environment  \textbf{19},  100351 (2020)

\bibitem{Shi2019}
Shi, Y., Li, F., Song, W., Li, X.Y., Ye, J.: Energy audition based cyber-physical attack detection system in iot. In: Proceedings of the ACM Turing Celebration ConferenceChina. ACM TURC '19, Association for Computing Machinery, New York, NY, USA (2019). \doi{10.1145/3321408.3321588}, \url{https://doi.org/10.1145/3321408.3321588}

\bibitem{sihwail2019malware}
Sihwail, R., Omar, K., Zainol~Ariffin, K.A., Al~Afghani, S.: Malware detection approach based on artifacts in memory image and dynamic analysis. Applied Sciences  \textbf{9}(18), ~3680 (2019)

\bibitem{tsoukas2022enhancing}
Tsoukas, V., Gkogkidis, A., Kampa, A., Spathoulas, G., Kakarountas, A.: Enhancing food supply chain security through the use of blockchain and tinyml. Information  \textbf{13}(5), ~213 (2022)

\end{thebibliography}

\end{document}